\documentclass[aps,prl,reprint,groupedaddress]{revtex4-1}

\usepackage{bm}
\usepackage{wasysym}
\usepackage{amssymb}
\usepackage[cp1251]{inputenc}
\usepackage[dvips]{graphicx}
\begin{document}

\title{Edge Plasmon Polaritons on a Half-Plane\\}

\author{A.A. Zabolotnykh}
\email{andrey.zabolotnyh@phystech.edu}
\affiliation{Kotelnikov Institute of Radio-engineering and Electronics of the RAS, Mokhovaya 11-7, Moscow 125009, Russia\\
}
\affiliation{Moscow Institute of Physics and Technology, Institutskii per. 9, Dolgoprudny, Moscow region 141700, Russia\\}

\author{V.A. Volkov}
\affiliation{Kotelnikov Institute of Radio-engineering and Electronics of the RAS, Mokhovaya 11-7, Moscow 125009, Russia\\
}
\affiliation{Moscow Institute of Physics and Technology, Institutskii per. 9, Dolgoprudny, Moscow region 141700, Russia\\}

\date{\today}

\begin{abstract}
The effect of electromagnetic retardation on the spectrum of edge plasmons in a semi-infinite two-dimensional
electron system is considered. The problem is reduced to complicated integral equations for the potentials,
which are solved upon a major simplification of the kernel. The spatial distribution of the potentials,
charges, and currents is analyzed. It is shown that edge plasmon polaritons in the high-conductivity two dimensional
system are characterized by a high Q factor at all frequencies, including those lower than the
inverse electron relaxation time $\tau^{-1}$.
\end{abstract}

\pacs{}
\maketitle

\section{Introduction}
Plasmons in two-dimensional (2D) electron systems
formed in semiconductor heterostructures have
been investigated for almost half a century, beginning
with the pioneering theoretical work \cite{Stern} and first
experimental studies \cite{Grimes,Allen,Theis}.

In contrast to the three-dimensional case, the
spectrum of 2D plasmons has no gap at zero wave vector
\cite{Stern,Chaplik} and, for a clean system (i.e., for the electron
relaxation time $\tau\to \infty$), in the quasistatic approximation
and in the long-wavelength limiting case can be
written as $\omega_0(q)=\sqrt{2\pi n_0 e^2q/\varkappa m}$, where $q$ is the magnitude
of the plasmon wave vector, $n_0$ is the unperturbed
2D electron density in the system, $m$ is the
effective mass of the electron, $e>0$ is the elementary
charge, and $\varkappa$ is the background dielectric constant.

For a finite electron relaxation time $\tau$, the plasmon
frequency becomes complex-valued, with its imaginary
part describing the plasmon decay with time. In
the above approximations, the dispersion relation can
be written as

\begin{equation}
	\label{spectr}
	\omega_{bulk}(q)=\omega'+i\omega''=\sqrt{\omega_0^2(q)-1/4\tau^2}-i/2\tau.
\end{equation}

One can see that the plasmon has a high Q factor (i.e.,
$\omega'\gtrsim\omega''$) in the frequency range above approximately
$1/\tau$. At lower frequencies, the plasmon decays more
rapidly than it oscillates.

However, the situation changes radically if electromagnetic
retardation is taken into consideration \cite{Govorov,Falko}.
The properties of 2D plasmon polaritons, i.e., plasmons
with electromagnetic retardation taken into
account, in a 2D electron system with a finite relaxation
time $\tau$, were analyzed in \cite{Falko}. It was found that the
spectrum depends considerably on the dimensionless
conductivity $\widetilde{\sigma}=2\pi\sigma/c$, where $\sigma=e^2n_0\tau/m$ is the
static conductivity of the 2D electron system (which
has the speed dimensionality in Gaussian units) and $c$ is
the speed of light (for simplicity, we consider a 2D system
in vacuum; i.e., $\varkappa=1$). If the conductivity of the
2D system is low ($\widetilde{\sigma}<1$), plasmon polaritons have a
low Q factor and their spectrum is qualitatively similar
to the plasmon spectrum given by Eq. (\ref{spectr}) in the sense
that the condition $\omega'\geq\omega''$ is only satisfied beginning
with some finite frequency $\omega'$ and wave vector $q$.
However, if the conductivity of the 2D system is high
($\widetilde{\sigma}>1$), the plasmon polariton spectrum changes significantly
(see \cite{Falko}, Fig. 1). In such a system, plasmon
polaritons have a high Q factor, i.e., $\omega'\gtrsim\omega''$, for all values
of the wave vector $q$ and frequency $\omega'$, including
frequencies $\omega'< 1/\tau$.

It is known that edge plasmons traveling along the
boundary of a 2D system can exist \cite{Mast, Glattli, Volkov1985, Fetter, Volkov1988, Volkov_Galch}. In the quasistatic
limiting case, their dispersion relation $\omega_{edge}(q)$
is similar to that of bulk plasmons (plasmons in a system
with no boundary), see Eq. (\ref{spectr}):
\begin{equation}
	\label{spectr_edge}
	\omega_{edge}(q)=\sqrt{\alpha^2\omega_0^2(q)-1/4\tau^2}-i/2\tau,
\end{equation}
where $q$ is the wave vector along the boundary and the
constant $\alpha\approx 0.906$ according to the exact solution of
the problem given in \cite{Volkov1985} and $\alpha=\sqrt{2/3}\approx 0.816$
according to the approximate solution given in \cite{Fetter}.
For $\omega'\tau < 1$, edge plasmons are strongly damped, similarly
to bulk plasmons in the quasistatic limiting case.

The importance of electromagnetic retardation
effects was noted in experimental studies on 2D plasmons \cite{Kukushkin_PRL2003, Kukushkin_JETPL2003, Kukushkin2006, Muravev2011, Gusikhin2014, Muravev2015}. We should also mention theoretical
studies on the microwave response of antidot arrays
\cite{Mikhailov1996} and stripe-shaped 2D electron systems \cite{Mikhailov2005}. The
spectrum of plasmon polaritons in a double-layer system
with a finite $\tau$ was considered in \cite{Chaplik2015}.

The goal of this study is to analyze the impact of
electromagnetic retardation on the spectrum of edge
plasmons (i.e., edge plasmon polaritons) in the simplest
semi-infinite 2D electron system, which is a half-plane.
It is essential that we use the simplest approach,
developed by Fetter \cite{Fetter} in his treatment of the edge plasmon
spectrum disregarding retardation.

\section{MAIN EQUATIONS AND THE METHOD
OF SOLUTION}

Let us consider a 2D electron system in vacuum
($\varkappa=1$) occupying the half-plane $x>0$, $z=0$, so that
its boundary coincides with the $y$-axis. We assume that
the electron density drops stepwise to zero for $x<0$.

We are going to determine the spectrum of edge
plasmon polaritons in the long-wavelength limiting
case $q\ll k_F$ (where $\hbar k_F$ is the Fermi momentum),
because retardation plays the most important role
when the plasmon wavelength is on the order of the
wavelength of light with the same frequency. We will
use the classical equation for the average velocity of
electrons (the Euler equation) and Maxwell’s equations.

The equation for the average electron velocity ${\bf v}= (v_x,v_y)$
can be written as follows (see, e.g., \cite{Fetter,Rudin}):
\begin{equation}
\label{hydro}
\partial_t {\bf v}+{\bf v}/\tau=-s^2\nabla n/n_0-e{\bf E}/m,
\end{equation}
where ${\bf E}=-{\bf\nabla} \varphi-\partial_t {\bf A}/c$ is the mean field induced by
electrons, ${\bf A}=(A_x,A_y)$ is the vector potential in the
plane of the 2D electron system, and $n$ is the perturbation
of the electron density with respect to its equilibrium
value. In the first term on the right-hand side of
Eq. (\ref{hydro}), which describes pressure, $s$ is about the Fermi
velocity $v_F$; $s^2=3v_F^2/4$ according to \cite{Rudin}.

Maxwell’s equations for the scalar potential $\varphi$ and
vector potential ${\bf A}=(A_x,A_y)$ and $A_z$ in the Lorentz
gauge have the form
\begin{eqnarray}
\label{system}
&&
\left(\frac{1}{c^2}\frac{\partial^2}{\partial t^2} -\Delta  \right)
 \left(\begin{array}{lll}
 \varphi \\ {\bf A} \\A_z
 \end{array}
 \right)=4\pi\left(\begin{array}{lll}
  \rho \\ {\bf j}/c \\ 0
  \end{array}
  \right)\delta(z), \\
 && div{\bf A}+\partial_zA_z+\partial_t\varphi/c=0, \nonumber
\end{eqnarray}
where $div=(\partial_x,\partial_y)$, $\rho=-en$ is the electron-density
perturbation, and ${\bf j}$ is the 2D current density in the 2D
electron system. A consequence of Eqs. (\ref{system}) is the continuity
equation $\partial_t\rho+div {\bf j}=0$.

In the case of an unbounded system, Eqs. (\ref{hydro}) and (\ref{system}) define the spectra of independent TE and TM
modes \cite{Falko}. For large wave vectors (away from the light
cone), the TM mode represents the conventional
longitudinal plasmon with the spectrum given by
Eq. (\ref{spectr}). The existence of a boundary leads to the mixing
of TE and TM modes.

Let us seek a solution in the form of a wave traveling
along the boundary: ${\bf A}={\bf A}(x,z) \exp(iq_yy-i\omega t)$ and  $\varphi=\varphi(x,z) \exp(iq_yy-i\omega t)$, where ${\bf A}(x,z)$ and $\varphi(x,z)$ decrease away from the boundary. The vector potential
component $A_z$ equals zero, because current does
not flow perpendicularly to the 2D electron system.
The component $A_y$ can be excluded using the gauge
condition. Thus, we focus on the equations for $\varphi(x,z)$
and $A_x(x,z)$:
\begin{equation}
\label{A_phi}
\left(\partial_x^2+\partial_z^2-\beta^2 \right)
 \left(\begin{array}{ll}
 \varphi(x,z) \\  A_x(x,z)
 \end{array}
 \right)=4\pi e\left(\begin{array}{ll}
  n(x) \\ \frac{n_0v_x(x)}{c} 
  \end{array}
  \right)\delta(z),
\end{equation}
where we have linearized the current density ${\bf j}=-en_0{\bf v}$, $\beta=\sqrt{q_y^2-\omega^2/c^2}$
with Re$\beta>0$, and $v_x(x)=0$ and $n(x)=0$ for $x<0$.

Using the Green’s functions technique, one can
reduce the set of equations (\ref{A_phi}) in 2D space to a
set of integral equations in one-dimensional (1D)
space. One can try to solve the resulting set of equations by the
Wiener–Hopf method (see, e.g., \cite{Volkov1985}). However, the
solution obtained in this way (if an explicit solution
can be derived at all) is very cumbersome and hard to
analyze. Thus, the plasmon spectrum in finite systems
is frequently obtained using approximate methods.
One of these methods, which we use here, is the simplification
of the kernel in the integral equation for $\varphi(x,z)$
(and $A_x(x,z)$).

Seemingly, the simplification of the kernel of the
integral equation was used for the first time to calculate
the dispersion relation of 2D plasmons in \cite{Fetter},
where the spectrum of the edge plasmon and magnetoplasmon
for a semi-infinite system with a straight
boundary was determined in the quasistatic approximation.
In the case of the edge plasmon (without an
external magnetic field), the spectrum obtained is in
good agreement with the one determined from the
exact solution \cite{Volkov1985}; the only difference is in the value
of the constant $\alpha$ (see Introduction). Thus, one can hope
that this method will yield reasonable results for the
spectrum of the edge plasmon polariton as well. We
mention that this method was used to calculate the
spectra of plasmons in a strip \cite{Cataudella}, edge magnetoplasmons
at the boundary between two 2D layers \cite{Mikhailov1995}, and
edge plasmons in graphene \cite{Wang} and topological systems \cite{Song,Kumar}.

Let us describe the essence of the method. Considering
the first equation of the set (\ref{A_phi}) (the second equation can be treated in a similar way) for $z=0$, i.e.,
within the plane of the 2D system, and let us transform it
to the 1D integral equation
\begin{equation}
\label{int_eq}
	\varphi(x,z=0)=-2 e\int_{-\infty}^{+\infty}K_0(\beta|x-x'|)n(x')dx'.
\end{equation}
Here, $K_0(x)$ is the zero-order modified Bessel function
of the second kind; its asymptotic behavior is $K_0(x)=\sqrt{2/(\pi x)}\exp(-|x|)$
for $x\to\infty$ and $K_0(x)=\ln(2/x)-\gamma$
for $x\to 0$, where $\gamma\approx 0.577$ is the
Euler-Mascheroni constant. The method consists in
the replacement of the kernel $K_0(x)$ with a simpler one $L_0(x)$
characterized by the same area under the curve
and the second moment (for details, see \cite{Fetter} and
references therein). It proves that $L_0(x)=\pi\exp(-\sqrt{2}\beta |x|)/\sqrt{2}$
can be taken as an approximation
for the kernel $K_0(x)$. As far as $L_0(x)$ is the Green’s
function for the operator $(-\partial_x^2+2\beta^2)/(2\pi\beta)$, we find
that, after replacing $K_0(x)$ with $L_0(x)$ in Eq. (\ref{int_eq}), the latter
can be transformed to the following set of differential
equations for $\varphi(x,0)$ and $A_x(x,0)$:
\begin{equation}
\label{simple_eq}
	\left\{ 
		\begin{array}{lcr}
			(\partial_x^2-2\beta^2) \varphi(x,0)=4\pi \beta en(x),\\
			(\partial_x^2-2\beta^2) A_x(x,0)=4\pi \beta e n_0v_x(x)/c,
			 \end{array}
	\right.
\end{equation}
where $n(x)$ and $v_x(x)$ are equal to zero for $x<0$ and
are finite for $x>0$. Thus, instead of integral equations
(\ref{A_phi}), we obtain differential equations (\ref{simple_eq}).

The substitution of the density $n$ from the continuity
equation and of the current $\bf{j}$ from Eq. (\ref{hydro}) into Eq. (\ref{simple_eq})
provides the following set of equations for $x>0$:
\begin{equation}
\label{system_fin}
\left\{
	\begin{array}{lcr}
   s^2\varphi''''(x)+(\omega\widetilde\omega-2\omega_p^2-2\beta^2s^2-s^2q_y^2)\varphi''(x)\\
   +2\beta^2(\omega_p^2-\omega\widetilde\omega+s^2q_y^2)\varphi(x)=0,\\
  A_x''(x)- 2(\beta^2+\omega_p^2\omega/(\widetilde{\omega}c^2))A_x(x)=\\ (2\omega_p^2\varphi'(x)+2\beta^2s^2\varphi'(x)-
  s^2\varphi'''(x))/(-i\widetilde{\omega}c),
	\end{array}
		\right.
\end{equation}
where $\widetilde\omega=\omega+i/\tau$, $\omega_p^2=2\pi e^2 n_0 \beta/m$, and $\varphi(x)$ and $A_x(x)$
are taken at $z=0$. We note that the equation for $\varphi(x)$
does not include the vector potential components.
For $x<0$, we evidently obtain the following
simple set of equations:
\begin{equation}
\label{system_fin_ad}
\left\{
	\begin{array}{lcr}
  (\partial_x^2-2\beta^2) \varphi(x)=0,\\
  (\partial_x^2-2\beta^2) A_x(x)=0.
 \end{array}
 		\right.
\end{equation}

Let us discuss the boundary conditions for Eqs. (\ref{system_fin}) and (\ref{system_fin_ad}). First, we seek solutions localized near the
boundary, i.e., decreasing for $x\to \pm \infty$. Second, we
assume that $\varphi(x)$ and $A_x(x)$ along with their first
derivatives are continuous at $x=0$; this follows from
Eq. (\ref{simple_eq}) and the absence of $\delta$-like (or even more
singular) distributions of charges and currents at the boundary \cite{edge}. Third, the current (or velocity) component
perpendicular to the boundary has to vanish at
the boundary: $v_x(x=0)=0$.

Now, we proceed to the solution of sets of
equations (\ref{system_fin}) and (\ref{system_fin_ad}). Solving first the equation for $\varphi(x)$, we obtain
\begin{eqnarray}
\label{phi}
	 &\varphi(x)=\varphi_0e^{\sqrt{2}\beta x}, &\quad x<0;\\
  &\varphi(x)=\varphi_1e^{-\lambda_1x}+ \varphi_2e^{-\lambda_2x},&\quad x>0;
 \end{eqnarray}
where $\varphi_{0,1,2}$ are constants. Provided that $s$ and $q_y$ are small, i.e. $s/c\ll 1$, $|\omega_p^2-\omega\widetilde{\omega}|\gg s^2q_y^2$, and $|2\omega_p^2-\omega\widetilde{\omega}|\gg s^2|q_y^2+2\beta^2|$, we obtain
\begin{equation}
 	\lambda_1^2=2\beta^2\frac{\omega_p^2-\omega\widetilde\omega}{2\omega_p^2-\omega\widetilde\omega},\quad \lambda_2^2=\frac{2\omega_p^2-\omega\widetilde\omega}{s^2}, 
\end{equation}
where the sign of $\lambda_{1,2}$ is determined from the condition
Re$\lambda_{1,2}>0$.

Next, we substitute the solution obtained for $\varphi(x)$
at $x>0$ into the right-hand side of the second equation
of the set (\ref{system_fin}). We obtain
\begin{eqnarray}
\label{A_x}
  & A_x(x)=A_0e^{\sqrt{2}\beta x},& \quad x<0;\,\\
  & A_x(x)=A_1e^{-\lambda_1x}+  A_2e^{-\lambda_2x}+A_3e^{-\gamma x},&\quad x>0;\,   
\end{eqnarray}
where $A_{0,1,2,3}$ are constants, $\gamma^2=2(\beta^2+\omega_p^2\omega/(\widetilde{\omega}c^2))$,
and Re$\gamma>0$. Constants $A_{1,2}$ are unambiguously
related to $\varphi_{1,2}$:
\begin{equation}
	A_{1,2}=\frac{\varphi_{1,2}\lambda_{1,2}(2\omega_p^2+2\beta^2s^2-s^2\lambda_{1,2}^2)}{i\widetilde{\omega}c(\lambda_{1,2}^2-\gamma^2)}.
\end{equation}
We have obtained five unknown constants ($\varphi_{0,1,2}$ and $A_{0,3}$) and five boundary conditions at $x=0$ (the
continuity of $\varphi(x)$, $A_x(x)$, and their first-order derivatives; and
the vanishing of the current at the boundary). Substituting
the expressions obtained for $\varphi(x)$ and $A_x(x)$ into the
boundary conditions, one can find the dispersion relation.
Under the condition that $\omega\widetilde{\omega}$ is of the same order
of magnitude as $2\omega_p^2-\omega\widetilde{\omega}$, along with the conditions
used to simplify the expressions for $\lambda_{1,2}$, we obtain
\begin{eqnarray}
\label{disp}
	\omega\widetilde{\omega}= D\left(2\omega_p^2(1-\delta)-\omega\widetilde{\omega}\right), \qquad \\
	\text{where}\, D=\sqrt{\frac{\omega_p^2-\omega\widetilde{\omega}}{2\omega_p^2-\omega\widetilde{\omega}}},
	\delta= \frac{\sqrt{1+(\omega_p^2\omega)/(c^2\widetilde{\omega}\beta^2)}-1}{D+\sqrt{1+(\omega_p^2\omega)/(c^2\widetilde{\omega}\beta^2)}}
	.\nonumber
\end{eqnarray}

In the quasistatic limiting case ($c\to \infty$), the coefficient $\delta$
tends to zero, and we obtain the dispersion
relation found in \cite{Fetter}. However, it is insufficient to
simply replace $q_y$ with $\sqrt{q_y^2-\omega^2/c^2}$ in the dispersion
relation without retardation in order to obtain
Eq. (\ref{disp}).

\section{ANALYSIS OF THE DISPERSION RELATION}

\begin{figure}
			\includegraphics[width=8cm]{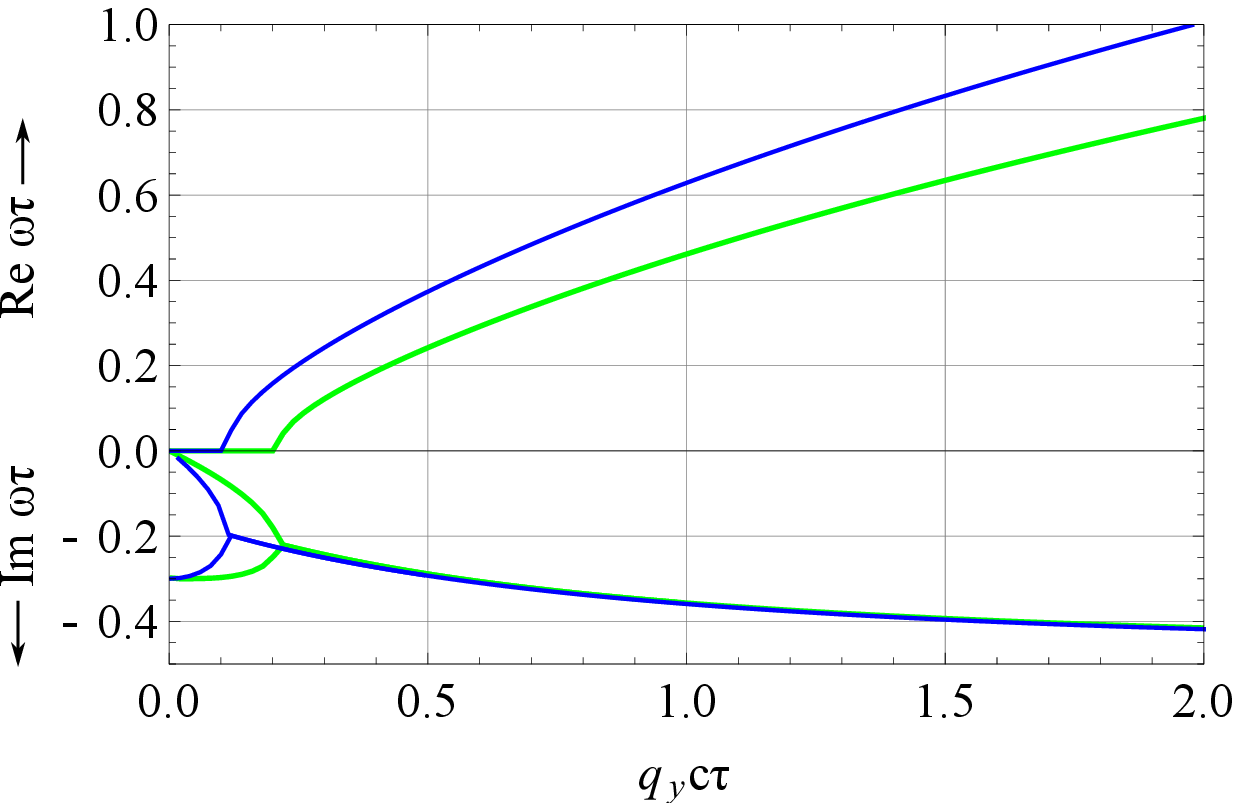}
			\caption{ \label{Fig:low}
			Spectra of plasmon polaritons in a
			2D system with the dimensionless conductivity
			$\widetilde{\sigma}=2\pi e^2n_0\tau/(mc)=0.7$. Green lines show the spectrum
			of the edge plasmon polariton; blue lines show the spectrum
			of the bulk plasmon polariton with the same value of $q_y$		and $q_x=0$. The real and imaginary parts of the complex-valued plasmon frequency, normalized to the electron
			relaxation time $\tau$, are given on the vertical axis in the top
			and bottom panels, respectively.
				}
\end{figure}

\begin{figure}
			\includegraphics[width=8cm]{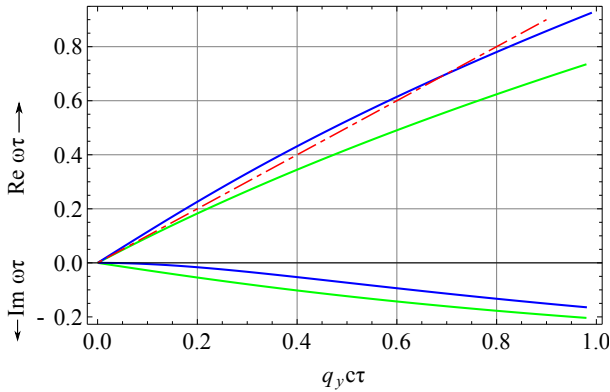}
			\caption{ \label{Fig:high}
			Spectra of plasmon polaritons in a
			2D system with the dimensionless conductivity $\widetilde{\sigma}=2$.
			Green lines show the spectrum of the edge plasmon polariton;
			blue lines show the spectrum of the bulk plasmon
			polariton with the same value of $q_y$ and $q_x=0$. The real
			and imaginary parts of the complex-valued plasmon frequency,
			normalized to the electron relaxation time $\tau$, are
			given on the vertical axis in the top and bottom panels,
			respectively. The red dash-dotted line shows the light cone ($\omega=cq_y$).
				}
\end{figure}

The dispersion relation represented by Eq. (\ref{disp})
makes it possible to determine the spectrum of edge
plasmon-polaritons, shown in Figs. \ref{Fig:low} and \ref{Fig:high}.

Let us first consider the case $\widetilde{\sigma}<1$ (Fig.~\ref{Fig:low}). In this case, at small wave vectors, plasmon polaritons are
characterized by a pure relaxation spectrum; for $q_y= 0$,
$\omega\tau=-i(1-\widetilde{\sigma})$ or $\omega=0$. For large wave vectors,
we obtain the usual spectrum given by Eq. (\ref{spectr_edge})
with $\alpha=\sqrt{2/3}$.

As $\widetilde{\sigma}$ approaches unity from below, the pure relaxation
region “contracts” and vanishes for $\widetilde{\sigma}=1$.

\begin{figure}
			\includegraphics[width=7.5cm]{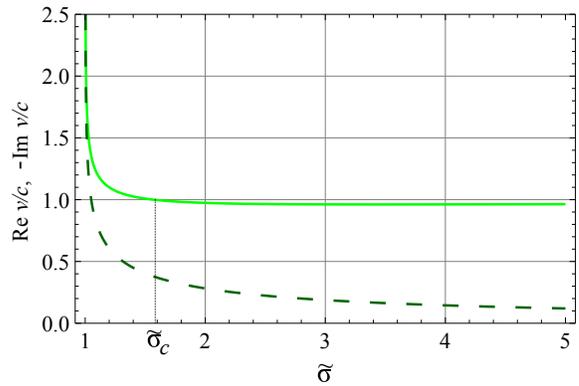}
			\caption{ \label{Fig:vel}
			(Solid line) Real part Re$v$ and
			(dashed line) imaginary part Im$v$ of the edge plasmon
			polariton velocity $v$ for $q_y\to 0$ versus the dimensionless
			conductivity $\widetilde{\sigma}>1$. 
				}
\end{figure}

The typical spectra of edge and bulk plasmon polariton
for $\widetilde{\sigma}>1$ are shown in Fig.~\ref{Fig:high}. Let us recall the characteristics of bulk plasmon polaritons for $\widetilde{\sigma}>1$. The
asymptotic behavior of its spectrum for $q\to 0$ can be
written as $\omega=\widetilde{\sigma}cq/\sqrt{\widetilde{\sigma}^2-1}- i\tau(\widetilde{\sigma}cq/(\widetilde{\sigma}^2-1))^2$. Thus, $\omega'\propto q$, $\omega''\propto q^2$, and plasmon polaritons are
high-Q excitations at small wave vectors. Turning now
to the spectrum of edge plasmon polaritons, we see
that the asymptotic behavior at $q\to 0$ is linear for both $\omega'$ and $\omega''$, in contrast to the bulk case; i.e., $\omega=v q_y$,
where $v$ is the complex-valued velocity. The $v(\widetilde{\sigma})$
dependence is plotted in Fig.~\ref{Fig:vel}. It exhibits the asymptotic
behavior
\begin{equation}
	v/c=
\left\{
	\begin{array}{lcr}
  \sqrt[4]{3}(1-i)/(2\sqrt[4]{\widetilde{\sigma}^2-1}), \quad \widetilde{\sigma}\to 1+0;\\
  1-(1+i\sqrt{3})/(4\sqrt[3]{4}\widetilde{\sigma}^{2/3}), \quad \widetilde{\sigma}\to\infty.
 \end{array}
 		\right.
\end{equation}

The real part of the velocity Re$v$ exceeds $c$ for
$1<\widetilde{\sigma}<\widetilde{\sigma}_c$, where $\widetilde{\sigma}_c\approx 1.57$ (see Fig.~\ref{Fig:vel}). For $\widetilde{\sigma}\to \infty$,
the real part of the velocity $v$ approaches $c$ from below,
while its imaginary part tends to zero, and $Re\,v > Im\,v$ for any $\widetilde{\sigma}>1$. In other words, edge plasmon polaritons
are $"$high-Q$"$ excitations ($\omega' > \omega''$) for arbitrarily small
values of $q_y$ and, consequently, for arbitrarily low frequencies $\omega'$.

\begin{figure}
			\includegraphics[width=7.5cm]{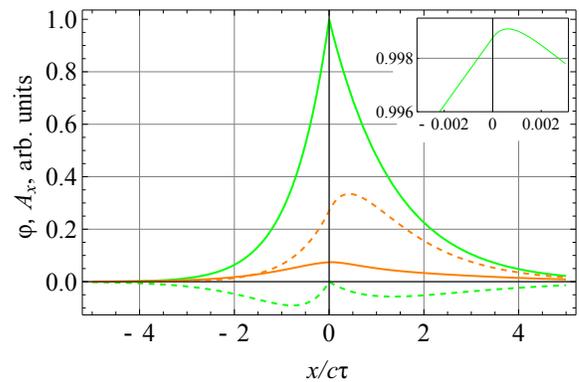}
			\caption{ \label{Fig:distrib}
			Complex amplitudes of the scalar
			potential $\varphi(x,z=0)$ and vector potential $A_x(x,z=0)$
			(green and orange lines, respectively) in relative units
			(assuming $\varphi_1=1$) versus the coordinate $x$ normalized by
			$c\tau$; the 2D electron system occupies the region $x>0$. The
			solid and dashed lines show the real and imaginary parts of
			the potentials, respectively. The inset shows the part of the $\varphi$
			dependence near zero on an expanded scale. The
			dependences are plotted for $\widetilde{\sigma}=2$, $q_yc\tau=1.2$, and $\omega\tau\approx 0.86-i 0.23$	(see Fig. \ref{Fig:high}). For the ratio $c/s$, a typical
			value of $10^3$ was assumed. The characteristic localization
			length is $1/Re\lambda_1\approx 1.4 c\tau $.
				}
\end{figure}

\begin{figure}
			\includegraphics[width=7.5cm]{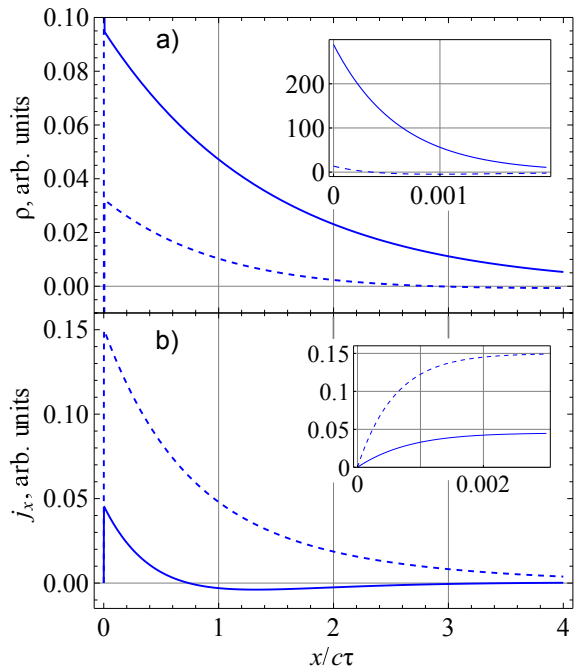}
			\caption{ \label{Fig:charge}
				Complex amplitudes of (a) the
						charge density $\rho(x)$ (normalized by $\varphi_1/c\tau$) and (b) the current
						density $j_x(x)$ (normalized by $\varphi_1/\tau$) versus the coordinate $x$
						normalized by $c\tau$. The solid and dashed lines show
						the real and imaginary parts, respectively, of $\rho$ and $j_x$. The insets show the parts of the corresponding
						dependences near zero on an expanded scale. The dependences
						are plotted for the same parameters as in Fig.~\ref{Fig:distrib}.	
				}
\end{figure}

The characteristic localization length of the edge
plasmon polariton field is determined by the root $\lambda_1$ and equals $1/Re\lambda_1$. For large values of $q_y$ (away from
the light cone), $Re\lambda_1\approx q_y/\sqrt{2}$. For $q_y\to 0$, we have $\lambda_1=q_y$ for $\widetilde{\sigma}\to 1+0$ and $\lambda_1\propto 1/\sqrt[3]{\widetilde{\sigma}}$ for $\widetilde{\sigma}\to \infty$.
Thus, for small $q_y$, the higher the conductivity, the
larger the plasmon localization region. For large $q_y$,
the localization region is independent of the conductivity
and is of the order of $q_y^{-1}$. The typical dependences
of the potentials $\varphi(x)$ and $A_x(x)$, as well as the
charge density $\rho(x)$ and current density $j_x(x)$, on the
coordinate $x$ are shown in Figs. \ref{Fig:distrib} and \ref{Fig:charge}. The parameters
of the system are given in the caption of Fig. \ref{Fig:distrib}.

Let us discuss the spatial distribution of the charge.
Some fraction of the charge is accumulated over a
short length $1/Re\lambda_2$ (see Fig.~\ref{Fig:charge}a, inset), while the rest
of it is distributed over a longer length $1/Re\lambda_1$. We estimated
the amount of charge concentrated at these two
length scales. For large wave vectors $q_y$ (when the
retardation effects are insignificant), approximately
the same amount of charge is distributed at these two
length scales. For $q_y\to 0$ and $\widetilde{\sigma}\to 1+0$, most of the charge is distributed at the longer length scale $1/Re\lambda_1$.
For $q_y\to 0$ and $\widetilde{\sigma}\to \infty$, about $30\%$ of all charge is distributed over the length $1/Re\lambda_1$.

\section{Conclusions}
To summarize, using the method developed by
Fetter \cite{Fetter}, we have found the approximate spectrum
of edge plasmon polaritons in a semi-infinite 2D system
with straight boundary. We have analyzed the spatial
distribution of the potentials, charges, and currents.
We have shown that, under the condition $2\pi\sigma>c$
(where $\sigma$ is the static conductivity of the 2D
system), the edge plasmon polariton, like its bulk
counterpart, can be a high-Q excitation (i.e., $Re\,\omega\gtrsim Im\,\omega$) for all wave vectors and, thus, for all frequencies,
including frequencies lower than $\tau^{-1}$, where $\tau$ is
the electron relaxation time in the 2D system.
\begin{acknowledgments}
The work was partly supported by the Russian Foundation for Basic Research (Projects Nos. 14-02-01166 and 16-32-00526). A.A.Z. acknowledges the support of
the Dynasty Foundation.
\end{acknowledgments}


\begin{thebibliography}{50} 
\bibitem{Stern}
F. Stern, Phys. Rev. Lett. {\bf 18}, 546 (1967).

\bibitem{Grimes}
C.\,C. Grimes and G. Adams, Phys. Rev. Lett. {\bf 36}, 145 (1976).

\bibitem{Allen}
S.\,J. Allen, Jr., D.\,C. Tsui, and R.\,A. Logan, Phys. Rev. Lett. {\bf 38}, 980 (1977).

\bibitem{Theis}
T.\,N. Theis, J.\,P. Kotthaus, and P.\,J. Stiles, Solid State Commun. {\bf 26}, 603 (1978).

\bibitem{Chaplik}
A.\,V. Chaplik, Sov. Phys. JETP {\bf 35}, 395 (1972).

\bibitem{Govorov}
A.\,O. Govorov and A.\,V. Chaplik, Sov. Phys. JETP {\bf 68},
1143 (1989).

\bibitem{Falko}
V.\,I. Fal’ko and D.\,E. Khmel’nitskii, Sov. Phys. JETP
{\bf 68}, 1150 (1989).

\bibitem{Mast}
D.\,B. Mast, A.\,J. Dahm, and A.\,L. Fetter, Phys. Rev. Lett. {\bf 54}, 1706 (1985).

\bibitem{Glattli}
D.\,C. Glattli, E.\,Y. Andrei, G. Deville, J. Poitrenaud, and F.\,I.\,B. Williams,
Phys. Rev. Lett. {\bf 54}, 1710 (1985).

\bibitem{Volkov1985}
V.\,A. Volkov and S.\,A. Mikhailov, JETP Lett. {\bf 42}, 556
(1985).

\bibitem{Fetter}
A.\,L. Fetter, Phys. Rev. B {\bf 32}, 7676 (1985).

\bibitem{Volkov1988}
V.\,A. Volkov and S.\,A. Mikhailov, Sov. Phys. JETP {\bf 67},
1639 (1988).

\bibitem{Volkov_Galch}
V.\,A. Volkov, D.\,V. Galchenkov, L.\,A. Galchenkov,
I.\,M. Grodnenskii, O.\,R. Matov, and S.\,A. Mikhailov,
JETP Lett. {\bf 44}, 655 (1986).

\bibitem{Kukushkin_PRL2003}
I.\,V. Kukushkin, J.\,H. Smet, S.\,A. Mikhailov, D.\,V. Kulakovskii, K. von Klitzing, and W. Wegscheider, Phys. Rev. Lett. {\bf 90}, 156801 (2003).

\bibitem{Kukushkin_JETPL2003}
I.\,V. Kukushkin, D.\,V. Kulakovskii, S.\,A. Mikhailov,
Yu. Smet, and K. von Klitzing, JETP Lett. {\bf 77}, 497
(2003).

\bibitem{Kukushkin2006}
I.\,V. Kukushkin, V.\,M. Muravev, J.\,H. Smet, M. Hauser, W. Dietsche, and K. von Klitzing, Phys. Rev. B {\bf 73}, 113310 (2006).

\bibitem{Muravev2011}
V.\,M. Muravev, I.\,V. Andreev, I.\,V. Kukushkin, S. Schmult, and W. Dietsche, Phys. Rev. B {\bf 83}, 075309 (2011).

\bibitem{Gusikhin2014}
P.\,A. Gusikhin, V.\,M. Muravev, and I.\,V. Kukushkin,
JETP Lett. {\bf 100}, 648 (2015).

\bibitem{Muravev2015}
V.\,M. Muravev, P.\,A. Gusikhin, I.\,V. Andreev, and I.\,V. Kukushkin, Phys. Rev. Lett. {\bf 114}, 106805 (2015).

\bibitem{Mikhailov1996}
S.\,A. Mikhailov, Phys. Rev. B {\bf 54}, 10335 (1996).

\bibitem{Mikhailov2005}
S.\,A. Mikhailov and N.\,A. Savostianova, Phys. Rev. B {\bf 71}, 035320 (2005).

\bibitem{Chaplik2015}
A.\,V. Chaplik, JETP Lett. {\bf 101}, 545 (2015).

\bibitem{Rudin}
S. Rudin and M. Dyakonov, Phys. Rev. B {\bf 55}, 4684 (1997).

\bibitem{Cataudella}
V. Cataudella and G. Iadonisi, Phys. Rev. B {\bf 35}, 7443 (1987).

\bibitem{Mikhailov1995}
S.\,A. Mikhailov, JETP Lett. {\bf 61}, 418 (1995).

\bibitem{Wang}
W. Wang, J.\,M. Kinaret, and S.\,P. Apell, Phys. Rev. B {\bf 85}, 235444 (2012).

\bibitem{Song}
J.\,C.\,W. Song and M.\,S. Rudner, PNAS {\bf 113}, 4658-4663 (2016).

\bibitem{Kumar}
A. Kumar, A. Nemilentsau, K.\,H. Fung, G. Hanson, N.\,X. Fang, and T. Low, Phys. Rev. B {\bf 93}, 041413(R) (2016).

\bibitem{edge}
L. A. Vainshtein, Diffraction Theory and Factorization
Method (Sov. Radio, Moscow, 1966), p. 12 [in Russian].
\end{thebibliography}
\end{document}